\documentclass[superscriptaddress,twocolumn,showpacs,prb,amsmath,amssymb]{revtex4}% Physical Review B

\usepackage{graphicx,bm}

\begin{document}

\title{Linear conductance in Coulomb-blockade quantum dots
in the presence of interactions and spin} 

\author{Y. Alhassid}
\affiliation{Center for Theoretical Physics, Sloane Physics
  Laboratory, Yale University,  New Haven, Connecticut 06520} 
\author{T. Rupp}
\affiliation{Center for Theoretical Physics, Sloane Physics
  Laboratory, Yale University,  New Haven, Connecticut 06520} 
\author{A. Kaminski}
\affiliation{Condensed Matter Theory Center,
Department of Physics, University of Maryland, College Park, Maryland
20742-4111}
\author{L. I. Glazman} 
\affiliation{Theoretical Physics Institute, University of Minnesota,
  Minneapolis, MN 55455} 

\begin{abstract}
  We discuss the calculation of the linear conductance through a
  Coulomb-blockade quantum dot in the presence of interactions beyond
  the charging energy. In the limit where the temperature is large
  compared with a typical tunneling width, we use a rate-equations
  approach to describe the transitions between the corresponding
  many-body states. We discuss both the elastic and
  rapid-thermalization limits, where the rate of inelastic scattering
  in the dot is either small or large compared with the elastic
  transition rate, respectively. In the elastic limit, we find several
  cases where a closed solution for the conductance is possible,
  including the case of a constant exchange interaction. In the
  rapid-thermalization limit, a closed solution is possible in the
  general case. We show that the corresponding expressions for the
  linear conductance simplify for a Hamiltonian that is invariant
  under spin rotations.
\end{abstract}

\maketitle

\section{Introduction}
In the Coulomb-blockade regime of quantum dots, the conductance of the
dot exhibits peaks as a function of the gate voltage, where each peak
corresponds to the tunneling of one more electron into the dot. Of
particular interest are diffusive or ballistic chaotic dots, where the
mesoscopic fluctuations of the conductance peaks and their spacings
reflect the statistical nature of the ground-state wave function and
energy of the isolated dot.\cite{alhassid00,aleiner02} The simplest
model to describe Coulomb blockade is the constant-interaction (CI)
model, in which the electrons occupy single-particle levels in the dot
and the interaction is described by an electrostatic Coulomb energy
that is constant for a fixed number of electrons. In this model, the
conductance near a Coulomb-blockade peak at temperatures that are
large compared with a typical tunneling width can be derived in a
closed form using a master-equations
approach.\cite{glazman88,beenakker91} This regime of sequential
tunneling is relevant to most Coulomb-blockade experiments involving
weakly-coupled dots.

At sufficiently low temperatures, the conductance through the dot is
dominated by elastic processes. However, at finite temperature, the
electrons in the dot may undergo inelastic scattering processes
caused, for example, by electron-phonon or electron-electron
interactions. In general, one may consider two limiting cases,
depending on the relative magnitude of a typical elastic tunneling
width $\Gamma_{\rm el}$ of an electron into the dot, and a typical
inelastic scattering width $\Gamma_{\rm in}$ of electrons in the dot.
In the so-called elastic limit, $\Gamma_{\rm el} \gg \Gamma_{\rm in}$,
and inelastic scattering processes in the dot can be ignored. In the
opposite limit, $\Gamma_{\rm el} \ll \Gamma_{\rm in}$, inelastic
scattering occurs on such a short time scale that electrons in the dot
are effectively thermalized immediately after an electron tunnels in
or out of the dot.\cite{beenakker91} This limit will be called the
rapid-thermalization limit, which in the literature is also sometimes
referred to as the inelastic limit. Statistical properties of the
conductance peaks were studied within the CI model in both
elastic\cite{jalabert92,alhassid98} and
rapid-thermalization\cite{beenakker01,rupp02,held02a} limits, as well
as for intermediate situations.\cite{eisenberg02}

However, recent experiments in almost-isolated Cou\-lomb-blockade dots
clearly indicate that it is necessary to take into account
electron-electron interactions beyond the CI model.\cite{spacings}
For such interactions, the eigenstates of the dot are no longer
characterized by Slater determinants. An important question that
arises is how to calculate the linear conductance through such a dot.
In this work we provide an answer in the limit of sequential
tunneling. We make the general assumption that both the thermal energy
and the typical excitation energies in the dot are small compared with
the charging energy. This defines the regime of Coulomb blockade,
where in the vicinity of the $N+1$-st conductance peak only the
manifolds of states with $N$ and $N+1$ electrons in the dot contribute
to the conductance (the manifolds with an electron number different
from either $N$ and $N+1$ are pushed away by the charging energy).
Given the many-body eigenstates of the dot (with $N$ and $N+1$
electrons), their energies, and the transition widths between the
states in the $N$- and $(N+1)$-electron dot as an electron tunnels
into the dot, we discuss the calculation of the linear conductance.
This is done by introducing non-equilibrium probabilities of the dot
to be in its various many-body states. These probabilities satisfy a
set of rate equations describing the transitions between the many-body
eigenstates. We consider stationary solutions in the linear-response
approximation in both limits where inelastic scattering of electrons
in the dot is weak (elastic limit) and strong (rapid-thermalization
limit). In the elastic limit, the linearized rate equations cannot be
solved in closed form in the general case, but there are several
important cases where a closed solution exits. Such cases include
sufficiently low temperatures where only ground-state to ground-state,
or ground-state to excited-states transitions (of $N$ and $N+1$
electrons, respectively) contribute to the conductance. Another case
where the rate equations can be solved in closed form at any
temperature corresponds to a Hamiltonian for which the occupations of
the orbital single-particle states are good quantum numbers.  A
particularly important such case corresponds to the universal
Hamiltonian\cite{kurland00,aleiner02} in the limit of infinite
Thouless conductance. This Hamiltonian includes a constant exchange
interaction in addition to a constant charging energy.\cite{usaj01}
We also consider the rapid-thermalization limit, where we can always
obtain a closed form expression for the conductance.

Furthermore, when the Hamiltonian of the dot is spin-rotation
invariant (as is the case of the Coulomb interaction), the expressions
we find for the linear conductance can be
simplified by carrying out explicitly the sum over the magnetic
quantum numbers (in both the elastic and rapid-thermalization limits).

The outline of this paper is as follows: In
Section~\ref{section:model} we introduce the basic assumptions and
notation.  In Section~\ref{section:rate-equations} we discuss the rate
equations satisfied by the probabilities to find the dot in one of its
many-body eigenstates with $N$ or $N+1$ electrons. We assume the
elastic limit, where inelastic scattering of electrons in the dot is
negligible. In Section~\ref{section:linear-response} we use
linear-response theory to linearize these master equations in the
stationary limit.  A general expression for the linear conductance in
terms of the solution to the rate equations is derived in
Section~\ref{section:linear-conductance}.  We show that if detailed
balance is satisfied for each pair of many-body eigenstates [in the
$N$- and $(N+1)$-electron dots, respectively], then the master
equations and consequently the conductance can be solved in closed
form.  Furthermore, the corresponding expressions are shown to be
simplified for Hamiltonians that are invariant with respect to spin
rotations.  Physically relevant cases where such closed solutions
exist are discussed in Section~\ref{solvable-conductance}, and include
the case of the universal Hamiltonian in the limit of infinite
Thouless conductance.  In Section~\ref{section:inelastic-scattering}
we solve the rate equations and derive a closed form for the
conductance in the case where the inelastic scattering width is much
larger than the elastic width (rapid-thermalization limit).  Finally,
in Section~\ref{section:summary} we summarize the main results of this
paper.

\section{Model}
\label{section:model}

We consider an almost-isolated quantum dot described by a Hamiltonian
$\hat H$ that includes a one-body part and a two-body interaction.
The dot is weakly coupled to leads, and we assume the limit $kT \gg
\Gamma$, where $\Gamma$ is a typical transition width of an electron
from the leads into the dot. In this limit we can ignore the coherence
between the dot and the leads and use a rate-equations approach to
study the transport through the dot. Furthermore, we assume the
Coulomb energy $e^2/C$ to be much larger than the thermal energy
$kT$. In this limit, the linear conductance of the dot exhibits
Coulomb-blockade peaks.  When the gate voltage $V_{\rm g}$ is tuned in
the vicinity of a Coulomb-blockade peak, the number of electrons in
the dot can be either $N$ or $N+1$. All manifolds with an electron
number different from either $N$ or $N+1$ are pushed away by the
charging energy.

The rate equations have been solved in the CI model for a single
spin-degenerate level\cite{glazman88} and for any number of
single-particle levels.\cite{beenakker91} In the CI model, the
two-body interaction is modeled by a constant charging energy of an
object with capacitance $C$ and charge $N e$. Here, we derive and
solve the rate equations in the presence of interactions beyond the
charging energy, where the many-body states of the dots are no longer
Slater determinants.

We denote the many-body eigenstates of the dot with $N$ electrons by
$i$, and their respective energies by $\varepsilon_i^{(N)}$.
Similarly, the states of the dot with $N+1$ electrons are labeled as
$j$ and the corresponding energies are $\varepsilon_j^{(N+1)}$.  $M_N$
and $M_{N+1}$ denote, respectively, the total number of the $N$- and
$(N+1)$-electron states considered.

Of particular interest are cases where the dot's Hamiltonian is
spin-rotation invariant. This requires the absence of a spin-orbit
interaction such that the single-particle energies of the electrons in
the dot $\epsilon_{\lambda}$ are spin-degenerate. In addition, for
spin-rotation invariance to be valid, the matrix elements of the
two-body interaction may depend only on the orbital states but not on
the spin indices, as it is the case for the Coulomb interaction. We
then have
\begin{equation}\label{spin}
[{\hat H},{\bf{\hat S}}^2]=0 \quad \textrm{and} \quad [{\hat H},{\hat
S}_z]=0\;.
\end{equation}
The many-body eigenstates of the dot can be characterized by the good
quantum numbers of the spin $S$ and its projection $S_z=M$. The
remaining quantum numbers, in addition to the total number of
electrons $N$, will be labeled by $\alpha$. Thus the eigenstates of
the dot with $n$ electrons are given by $i = (\alpha, S, M)$, and
their respective energies $\varepsilon_{\alpha S}^{(N)}$ are
independent of $M$. The states of the dot with $N+1$ electrons are
similarly labeled by $j = (\alpha', S', M')$, and their corresponding
energies are $\varepsilon_{\alpha'S'}^{(N+1)}$.  An example of a
spin-rotation invariant Hamiltonian is the recently proposed universal
Hamiltonian.\cite{kurland00,aleiner02}

\section{Rate equations}
\label{section:rate-equations}
 
We assume that a potential difference $V$ is applied between the two
leads (source and drain) at temperature $T$. Fractions $\eta_{\rm l}$
and $\eta_{\rm r}$ of this potential difference, with 
\begin{equation}
\label{etasum}
\eta_{\rm l}+\eta_{\rm r} = 1\,,
\end{equation}
fall between the dot and the left lead, and the right lead and the
dot, respectively.  As a result a current $I$ flows through the dot
and various eigenstates of the dot can be occupied with different
probabilities. We denote the non-equilibrium probability for the dot
to be in a particular $n$-electron state $i$ by $P_i^{(N)}$. Since we
consider situations in which the dot can only be occupied by $N$ or
$N+1$ electrons, we require the normalization
\begin{equation}
\sum_i P_i^{(N)} + \sum_j P_j^{(N+1)} = 1\,.
\label{normalization}
\end{equation}
We further assume that energy is conserved in the tunneling between
the dot and the leads, neglecting virtual transitions that are of
higher order in the tunneling widths. Denoting the transition widths
between the $N$-electron state $i$ and the $(N+1)$-electron state $j$
involving an electron tunneling into the dot from the left (right)
lead by $\Gamma_{ij}^{\rm l}$ ($\Gamma_{ij}^{\rm r}$), we can write
the following rate equations for the probabilities $P_i^{(N)}$ and
$P_j^{(N+1)}$:
\begin{widetext}
\begin{subequations}
\label{rate_eqs}
\begin{eqnarray}
\frac{\partial P_i^{(N)}}{\partial t} &=& \sum_j P_j^{(N+1)} 
[(1-f_{ij}^{\rm l}) \Gamma_{ij}^{\rm l} + (1-f_{ij}^{\rm r}) 
\Gamma_{ij}^{\rm r}]%\nonumber\\
%& & 
- P_i^{(N)} \sum_j (f_{ij}^{\rm l} \Gamma_{ij}^{\rm l} 
+ f_{ij}^{\rm r} \Gamma_{ij}^{\rm r})\quad \textrm{for all } i\,,
\label{rate_eqn_i}\\
\frac{\partial P_j^{(N+1)}}{\partial t} &=& \sum_i P_i^{(N)} 
(f_{ij}^{\rm l} \Gamma_{ij}^{\rm l} + f_{ij}^{\rm r} 
\Gamma_{ij}^{\rm r}) %\nonumber\\
%& & 
- P_j^{(N+1)} \sum_j [(1-f_{ij}^{\rm l}) \Gamma_{ij}^{\rm l} 
+ (1-f_{ij}^{\rm r}) \Gamma_{ij}^{\rm r}]\quad \textrm{for all } j\,.
\label{rate_eqn_j}
\end{eqnarray}
\end{subequations}
\end{widetext}
The Fermi-Dirac function of an electron in the left (right) lead is
evaluated at a suitable energy that can be found from energy
conservation of the transition between states $i$ and $j$, and is
denoted by $f_{ij}^{\rm l}$ ($f_{ij}^{\rm r}$). In the presence of a
gate voltage $V_{\rm g}$, the energy of the $n$-electron dot in state
$i$ is given by $\varepsilon^{(N)}_i - N e \zeta V_{\rm g}$. Here
$\zeta\equiv C_{\rm g}/C$, with $C_{\rm g}$ being the dot capacitance
with respect to the gate, and $C$ being the full dot capacitance.
Thus the energy of an electron (relative to the Fermi energy
$\epsilon_{\rm F}$ in the leads) in the left (right) lead that tunnels
into an $N$-electron dot in state $i$ to form an $(N+1)$-electron dot
in state $j$ is given by
\begin{equation}\label{energy-diff}
\varepsilon_{ij} =
\varepsilon_j^{(N+1)} - \varepsilon_i^{(N)} -  \tilde\epsilon_{\rm F}\;,
\end{equation}
where $\tilde\epsilon_{\rm F}\equiv e\zeta V_{\rm g} + \epsilon_{\rm
F}$ is an effective Fermi energy. Taking into account the
bias-potential drop between the dot and each lead, we have
\begin{equation}
\label{etas}
f_{ij}^{\rm l} \equiv f(\varepsilon_{ij} + \eta_{\rm l} e V)
\quad \textrm{and}\quad
f_{ij}^{\rm r} \equiv f(\varepsilon_{ij} - \eta_{\rm r} e V)\;,
\end{equation}
where the Fermi-Dirac function is defined by $f(x)
= [1+\exp(\beta x)]^{-1}$ with $\beta=1/kT$.

We note that we can choose any values of $\eta_{\rm l}$ and $\eta_{\rm
r}$ satisfying Eq.~(\ref{etasum}), but that the final result for the
physical quantities (e.g.\ the conductance) must be independent of
this choice (see the Appendix for a proof).

\section{Linear response}
\label{section:linear-response}

To calculate the conductance, we are interested in finding stationary
solutions of Eqs.~(\ref{rate_eqs}), i.e. non-equilibrium probabilities
$P_i^{(N)}$ and $P_j^{(N+1)}$ that are independent in time,
\begin{subequations}
\begin{eqnarray}
\frac{\partial P_i^{(N)}}{\partial t} = 0 & \textrm{for all } i\,,
\\ 
\frac{\partial P_j^{(N+1)}}{\partial t} = 0 & \textrm{for all } j\,.
\end{eqnarray}
\end{subequations}
Furthermore, to calculate the linear conductance, we are interested in
calculating the current for small bias voltages $V$, where the
equations can be solved in linear-response theory. To that end, we
linearize Eqs.~(\ref{rate_eqs}) in $V$. This involves an expansion of
the Fermi-Dirac distribution functions $f_{ij}^{\rm l}$, $f_{ij}^{\rm
  r}$ as well as the probabilities $P_i^{(N)}$ and $P_j^{(N+1)}$
around their equilibrium values. The Fermi-Dirac functions are
expanded up to first order as
\begin{equation}
f^{\rm l}_{ij} = f_{ij} + \eta_{\rm l} eV f_{ij}' \quad {\rm and} \quad
f^{\rm r}_{ij} = f_{ij} - \eta_{\rm r} eV f_{ij}' \,,
\label{fermi_dirac_exp}
\end{equation}
where we used the short-hand notation $f_{ij} = f(\varepsilon_{ij})$.
The probabilities $P_i^{(N)}$ and $P_j^{(N+1)}$ are expanded around
their equilibrium values ${\tilde P}_i^{(N)}$ and ${\tilde
  P}_j^{(N+1)}$ which they would attain if the bias voltage would be
zero and the dot would be in thermal equilibrium,
\begin{subequations}
\label{prob_exp}
\begin{eqnarray}
P_i^{(N)} &=& {\tilde P}_i^{(N)}[1+eV \beta \Psi_i^{(N)}]\,,
\label{prob_i_n_exp}
\\
P_j^{(N+1)} &=& {\tilde P}_j^{(N+1)}[1+eV \beta \Psi_j^{(N+1)}]
\label{prob_j_np1_exp}\,,
\end{eqnarray}
\end{subequations}
in terms of new variables $\Psi_i^{(N)}$ and $\Psi_j^{(N+1)}$. The
equilibrium probabilities can be expressed explicitly in terms of the
eigenenergies $\varepsilon_i^{(N)}$ and $\varepsilon_j^{(N+1)}$ of the
dot with $N$ and $N+1$ electrons,
\begin{subequations}
\label{def_equi_prob}
\begin{equation}
{\tilde P}_i^{(N)} =
{e^{-\beta(\varepsilon_i^{(N)}- \tilde\epsilon_{\rm F}N)} \over Z} 
\end{equation}
and
\begin{equation}
{\tilde P}_j^{(N+1)} =
{e^{-\beta [\varepsilon_j^{(N+1)} - \tilde \epsilon_{\rm F}(N+1)]}\over Z} \,,
\end{equation}
\end{subequations}
with the partition sum
\begin{equation}\label{partition}
Z = \sum_i e^{-\beta(\varepsilon_i^{(N)}- \tilde\epsilon_{\rm F}N)}
 + \sum_j  e^{-\beta[\varepsilon_j^{(N+1)} - \tilde\epsilon_{\rm F}(N+1)]}\,.
\end{equation}
The equilibrium distribution is in principle grand-canonical with a
chemical potential equal to the Fermi energy $\epsilon_{\rm F}$ (at
temperature $T$) in the leads, but because of the presence of charging
energy only two values of the electron numbers are allowed.  To find a
set of equations for the new variables $\Psi_i^{(N)}$ and
$\Psi_j^{(N+1)}$, we substitute the expansions (\ref{fermi_dirac_exp})
and (\ref{prob_exp}) into the r.h.s.\ of the rate
equations~(\ref{rate_eqs}) and equate the results to zero to ensure
stationarity. Keeping only the terms linear in the bias voltage $V$,
we obtain
\begin{widetext}
\begin{subequations}
\begin{eqnarray}
\lefteqn{
\sum_j {\tilde P}_j^{(N+1)} [\beta \Psi_j^{(N+1)} 
(1-f_{ij})(\Gamma_{ij}^{\rm l} + \Gamma_{ij}^{\rm r}) - 
f_{ij}'(\eta_{\rm l} \Gamma_{ij}^{\rm l} - \eta_{\rm r} 
\Gamma_{ij}^{\rm r})]}
\nonumber\\
&& \hspace{30mm} = {\tilde P}_i^{(N)} [\beta \Psi_i^{(N)} \sum_j f_{ij} 
(\Gamma_{ij}^{\rm l} +
\Gamma_{ij}^{\rm r}) + \sum_j f_{ij}' (\eta_{\rm l} \Gamma_{ij}^{\rm l} -
\eta_{\rm r} \Gamma_{ij}^{\rm r})] 
%\nonumber\\
%&&\hspace{75mm} 
\quad \textrm{for all } i\,,
\end{eqnarray}
\begin{eqnarray}
\lefteqn{
\sum_i {\tilde P}_i^{(N)} [\beta \Psi_i^{(N)} f_{ij} 
(\Gamma_{ij}^{\rm l} +
\Gamma_{ij}^{\rm r}) - f_{ij}' (\eta_{\rm l} 
\Gamma_{ij}^{\rm l} - \eta_{\rm r}
\Gamma_{ij}^{\rm r})]} 
\nonumber\\
&& \hspace{30mm}
= {\tilde P}_j^{(N+1)} [\beta \Psi_j^{(N+1)} \sum_i (1-f_{ij}) 
(\Gamma_{ij}^{\rm l}
+ \Gamma_{ij}^{\rm r}) - \sum_i f_{ij}' (\eta_{\rm l} 
\Gamma_{ij}^{\rm l} -
\eta_{\rm r} \Gamma_{ij}^{\rm r})]
%\nonumber\\
%&&\hspace{85mm} 
\quad \textrm{for all } j\,.
\end{eqnarray}
\end{subequations}
The above equations can be simplified utilizing the following three
relations. From the definition (\ref{def_equi_prob}) of the
equilibrium probabilities, it follows that
\begin{equation}
{\tilde P}_j^{(N+1)} = {\tilde P}_i^{(N)} e^{-\beta \varepsilon_{ij}}\;,
\label{equi_prob_n_np1}
\end{equation}
while the Fermi-Dirac distribution satisfies
\begin{equation}
[1-f(x)] e^{-\beta x} = f(x)
\label{fermi_dirac_rel1}
\end{equation}
and
\begin{equation}
f'(x) = - \beta f(x) / [1+ \exp(-\beta x)]\,.
\label{fermi_dirac_rel2}
\end{equation}
Using Eqs.~(\ref{equi_prob_n_np1}), (\ref{fermi_dirac_rel1}), and
(\ref{fermi_dirac_rel2}), we arrive at the transformed rate equations
\begin{subequations}
\label{trafo_rate_eqs}
\begin{eqnarray}
\sum_j f_{ij} [(\Gamma_{ij}^{\rm l} + \Gamma_{ij}^{\rm r})
(\Psi_j^{(N+1)}\!-\!\Psi_i^{(N)}) + (\eta_{\rm l}
\Gamma_{ij}^{\rm l} - \eta_{\rm r} 
\Gamma_{ij}^{\rm r})]&=&0
\quad\textrm{for all } i\,,
\label{trafo_rate_eqns_1} 
\\ \sum_i (1-f_{ij})
[(\Gamma_{ij}^{\rm l} + \Gamma_{ij}^{\rm r}) (\Psi_j^{(N+1)}\!-\!
\Psi_i^{(N)}) + (\eta_{\rm l} \Gamma_{ij}^{\rm l} - \eta_{\rm r}
\Gamma_{ij}^{\rm r})]&=& 0 \quad\textrm{for all } j\,.
\label{trafo_rate_eqns_2}
\end{eqnarray}
\end{subequations}
\end{widetext}
Equations~(\ref{trafo_rate_eqs}) represent a system of $M_N + M_{N+1}$
linear equations for $M_N + M_{N+1}$ unknowns. However, only $M_N +
M_{N+1} -1$ of these equations are linearly independent. Hence, the
solutions for $\Psi_i^{(N)}$ and $\Psi_j^{(N+1)}$ are determined only
up to an additive constant. This constant is fixed by imposing the
constraint
\begin{equation}\label{normalization1}
\sum_i {\tilde P}_i^{(N)}    \Psi_i^{(N)}   + 
\sum_j {\tilde P}_j^{(N+1)} \Psi_j^{(N+1)} = 0\,,
\end{equation}
that follows from the normalization condition~(\ref{normalization}).
In general, it is necessary to solve Eqs.~(\ref{trafo_rate_eqs})
numerically. These equations represent sets of detailed-balance
equations. However, it is possible to find analytic solutions in cases
where a detailed-balance equation holds for each pair of states $i$
and $j$ individually, i.e.\ in cases where each term in the square
brackets of Eqs.~(\ref{trafo_rate_eqs}) vanishes. This condition is
equivalent to
\begin{equation}
\Psi_i^{(N)} - \Psi_j^{(N+1)} =
\frac{\eta_{\rm l} \Gamma_{ij}^{\rm l} - \eta_{\rm r}
\Gamma_{ij}^{\rm r}}{\Gamma_{ij}^{\rm l} + \Gamma_{ij}^{\rm r}}
\quad \textrm{for any }i \textrm{ and } j\;.
\label{individual_detailed_balance}
\end{equation}
We emphasize that there are $M_N M_{N+1}$ equations
(\ref{individual_detailed_balance}) for only $M_N+M_{N+1}$ unknowns,
so in the general case it is not possible to satisfy
(\ref{individual_detailed_balance}). However, there are a number of
important cases where a solution does exist, as we discuss in
Section~\ref{solvable-conductance}.  A solution to
Eqs.~(\ref{individual_detailed_balance}), when it exists, does not
depend on temperature and leads to an expression for the conductance
in which the dependence on temperature and on decay widths factorizes.

\section{Linear conductance}
\label{section:linear-conductance}

The linear conductance $G$ is defined as the ratio $I/V$ in the limit
$V\to 0$.  The current through the dot can be expressed in terms of
the quantities considered in the master-equation approach. In this
framework, we can calculate for example the current $I$ through the
left tunneling contact which equals the current through the dot and
through the right contact,
\begin{equation}
I = \frac{e}{\hbar} \sum_{ij} [P_i^{(N)} f_{ij}^{\rm l}
- P_j^{(N+1)} (1-f_{ij}^{\rm l})]\Gamma_{ij}^{\rm l}\,.
\label{current}
\end{equation}
Substituting the expansions in Eqs.~(\ref{fermi_dirac_exp}) and
(\ref{prob_exp}) into Eq.~(\ref{current}), we find the elastic linear
conductance
\begin{equation}\label{elastic_conductance}
G_{\rm el} = \frac{e^2}{\hbar kT} \sum_{ij} {\tilde P}_i^{(N)} f_{ij} 
(\Psi_i^{(N)} - \Psi_j^{(N+1)} - \eta_{\rm l}) \Gamma_{ij}^{\rm l}\;.
\end{equation}
Although $\eta_{\rm l}$ appears explicitly in
Eq.~(\ref{elastic_conductance}), the linear conductance is independent
of $\eta_{\rm l}$. This must be so on physical grounds, and a formal
proof is provided in the Appendix.

In general, we have to solve Eqs.~({\ref{trafo_rate_eqs}) for
$\Psi_i^{(N)}$ and $\Psi_j^{(N+1)}$ and substitute their values in
Eq.~(\ref{elastic_conductance}). However, in the special case where
Eqs.~(\ref{individual_detailed_balance}) hold, the conductance is
given by the closed form
\begin{subequations}
\label{elastic-conductance}
\begin{equation}
G_{\rm el} = \frac{e^2}{\hbar kT} \sum_{ij} 
{\tilde P}_i^{(N)} f_{ij} \frac{\Gamma_{ij}^{\rm l}
\Gamma_{ij}^{\rm r}}{\Gamma_{ij}^{\rm l} + \Gamma_{ij}^{\rm r}} \;,
\end{equation}
or equivalently
\begin{equation}
G_{\rm el} = \frac{e^2}{\hbar kT} \sum_{ij} 
{\tilde P}_j^{(N+1)} (1-f_{ij}) \frac{\Gamma_{ij}^{\rm l}
\Gamma_{ij}^{\rm r}}{\Gamma_{ij}^{\rm l} + \Gamma_{ij}^{\rm r}}\,.
\label{elastic-conductance1}
\end{equation}
\end{subequations}
Notice that the parameters $\eta_{\rm l,r}$ no longer appear in the
conductance formula. The independence of the conductance from the
values of $\eta_{\rm l,r}$, which holds true in general (cf.\
Appendix), becomes apparent in this special case where an explicit
formula can be given. For the conductance in
Eqs.~(\ref{elastic-conductance}), the temperature dependence of the
contribution from each pair of states in the $N$-electron and
$(N+1)$-electron dots is independent from the transition widths.

So far, the spin symmetries have not been taken into account in the
derivation of the rate equations and the conductance. If this is done,
the sum over the many-body states $i$ and $j$ can be further
simplified. For two many-body states $i=(\alpha,S,M)$ and
$j=(\alpha',S',M')$, the associated partial decay widths
$\Gamma_{ij}^{\rm l}$ and $\Gamma_{ij}^{\rm r}$ are given by
\begin{equation} 
\Gamma_{ij}^{\rm l,r} = \Gamma_{0}^{\rm l,r} |\langle \alpha' S' M'|
\psi_m^\dag({\bf r}_{\rm l,r}) | \alpha S M \rangle |^2\,,
\label{decay_width}
\end{equation}
where the operator $\psi_m^\dag({\bf r}_{\rm l})$ [$\psi_m^\dag({\bf
r}_{\rm r})$] creates an electron with spin projection $m$ at the left
(right) point contact at ${\bf r}_{\rm l}$ (${\bf r}_{\rm r}$), and we
have introduced overall coupling strengths $\Gamma_0^{\rm l,r}$ at the
left and right point contact. The spin selection rules require
$m=M'-M$ and $S'=|S\pm 1/2|$ for the matrix element in
Eq.~(\ref{decay_width}) not to vanish. For a Hamiltonian which is
invariant with respect to spin rotations, we can use the Wigner-Eckart
theorem to factorize the matrix element in Eq.~(\ref{decay_width})
into a Clebsch-Gordan (CG) coefficient
%Clebsh-Gordan (CG) coefficient (or a Wigner 3$j$-symbol)
and a reduced matrix element that is independent of the
spin-projection quantum numbers $M$, $M'$, and $m$,
\begin{widetext}
\begin{eqnarray}\label{Wigner-Eckart}
\langle \alpha' S' M' | \psi_m^\dag({\bf r})
| \alpha S M \rangle &=&   -{1\over \sqrt{2S'+1}} (S \,M\, 1/2 \,m|S' \,M')
(\alpha' S' \| \psi^\dag({\rm \bf r}) \| \alpha S)
\nonumber\\
&=&  (-1)^{S'-M'} 
\left (
\begin{array}{c c c}
S' & 1/2 & S \\
M' & m & M
\end{array}
\right )
(\alpha' S' \| \psi^\dag({\rm \bf r}) \| \alpha S)\,.
\end{eqnarray}
For completeness, we have given the relation in
Eq.~(\ref{Wigner-Eckart}) both in terms of a CG coefficient (first
line) and a Wigner 3$j$-symbol (second line).  For a given pair of
many-body states $i$ and $j$, the decay widths $\Gamma_{ij}^{\rm l}$
and $\Gamma_{ij}^{\rm r}$ involve the same CG coefficient. In
Eq.~(\ref{individual_detailed_balance}), the CG coefficients in the
ratio on the r.h.s.\ cancel and we are left with
\begin{equation} 
 \Psi_i^{(N)}
 - \Psi_j^{(N+1)} = \frac{\eta_{\rm l}  \Gamma_0^{\rm l}|(\alpha' S' \|
\psi^\dag({\rm \bf r}_l) \| \alpha S)|^2 - \eta_{\rm r} 
\Gamma_0^{\rm r} |(\alpha' S'
\| \psi^\dag({\rm \bf r}_r)\| \alpha S)|^2}
{\Gamma_0^{\rm l} |(\alpha' S' \| \psi^\dag({\rm \bf r}_l) \| \alpha S)|^2 
+ \Gamma_0^{\rm r} |(\alpha' S' \| \psi^\dag({\rm \bf r}_r) \| 
\alpha S)|^2}\,.
\label{idb_simplified}
\end{equation}
\end{widetext}
Thus, if solutions for $\Psi_i^{(N)}$ and $\Psi_j^{(N+1)}$ exist that
satisfy Eq.~(\ref{individual_detailed_balance}), they will be
independent of the spin projection $M$ or $M'$, i.e.\ $\Psi_i^{(N)} =
\Psi_{\alpha S}^{(N)}$ and $\Psi_j^{(N+1)} = \Psi_{\alpha'
S'}^{(N+1)}$. Defining reduced decay widths
\begin{equation}\label{reduced-widths}
\tilde{\Gamma}_{\alpha S, \alpha' S'}^{\rm l,r} =
\Gamma_0^{\rm l,r} |(\alpha' S' \| \psi^\dag({\rm \bf r}_{\rm l,r}) \|
\alpha S)|^2 \;,
\end{equation}
conditions (\ref{idb_simplified}) can be rewritten in a form in which the
magnetic quantum numbers disappear,
\begin{equation}\label{idb_simplified1}
\Psi_{\alpha' S'}^{(N+1)} -
\Psi_{\alpha S}^{(N)} = \frac{\eta_{\rm l} \tilde{\Gamma}_{\alpha S,\alpha' 
S'}^{\rm l}
- \eta_{\rm r} \tilde{\Gamma}_{S,\alpha' S'}^{\rm r}} 
{\tilde{\Gamma}_{\alpha S,\alpha' S'}^{\rm l} + 
\tilde{\Gamma}_{\alpha S,\alpha' S'}^{\rm r}}\,.
\end{equation}
Using the reduced widths in the expression
 (\ref{elastic-conductance}) for the linear conductance, we have
\begin{widetext}
%\begin{eqnarray}
\begin{equation}
\label{closed-conductance-pre}
G_{\rm el} 
%& 
= 
%& 
\frac{e^2}{\hbar kT} \sum_{\alpha S \alpha' S'} \sum_{M M' m}
{\tilde P}_{\alpha S}^{(N)} f(\varepsilon_{\alpha S, \alpha' S'}) 
%\nonumber \\  && \times 
{1\over \ 2S'+1} |(S \,M\, 1/2 \,m|S' \,M')|^2
\frac{\tilde{\Gamma}_{\alpha S, \alpha' S'}^{\rm l}
\tilde{\Gamma}_{\alpha S, \alpha' S'}^{\rm r}} {\tilde{\Gamma}_{\alpha
S, \alpha' S'}^{\rm l} + \tilde{\Gamma}_{\alpha S, \alpha' S'}^{\rm r}}\,.
%\end{eqnarray}
\end{equation}
Taking advantage of the unitarity of the CG, $\sum_{M m}|(S \,M\, 1/2
\,m|S' \,M')|^2=1$, we can write as our final result for the linear
conductance
\begin{equation}\label{closed-conductance}
G_{\rm el} = \frac{e^2}{\hbar kT} \sum_{\alpha S \alpha' S'}
{\tilde P}_{\alpha S}^{(N)} f(\varepsilon_{\alpha' S', \alpha S})
\frac{\tilde{\Gamma}_{\alpha S, \alpha' S'}^{\rm l} 
\tilde{\Gamma}_{\alpha S, \alpha' S'}^{\rm r}} 
{\tilde{\Gamma}_{\alpha S, \alpha' S'}^{\rm l} + 
\tilde{\Gamma}_{\alpha S, \alpha' S'}^{\rm r}}\,.
\end{equation}
\end{widetext}
Instead of using the reduced widths defined in
Eq.~(\ref{reduced-widths}), it is possible to express the conductance
in terms of the widths for the maximally-projected spin states (i.e.\
$M=S$ and $M'=S'$)
\begin{equation}\label{maxproj}
\Gamma_{\alpha S, \alpha' S'}^{\rm
  l,r} \equiv \Gamma_{\alpha S S, \alpha' S' S'}^{\rm l,r}\;.
\end{equation}  
The reduced matrix elements in Eq.~(\ref{reduced-widths}) are non-zero
only for $S'=S\pm 1/2$ and, using the corresponding CG coefficients,
we have
\begin{equation}
\label{reduced-width-maxproj}
\tilde\Gamma_{\alpha S, \alpha' S'}^{\rm l,r} = 
[2\, {\rm max}(S,S') +1]\; \Gamma_{\alpha S, \alpha' S'}^{\rm l,r} \;.
\end{equation}

\section{Explicit solutions in the limit of elastic scattering}
\label{solvable-conductance} 

In general, it is not possible to find a closed solution for the
conductance in the limit of elastic scattering, and
Eqs.~(\ref{trafo_rate_eqs}) have to be solved numerically.  However,
in the following we list four important cases where explicit solutions
to Eqs.~(\ref{idb_simplified}) or more generally to
Eqs.~(\ref{trafo_rate_eqs}) exist and the conductance is given by a
closed expression.

\subsection{Ground-state transition at low temperatures}
\label{subsec:lowtemp}

At low enough temperatures, only the transition between the ground
states of the $n$- and $(N+1)$-electron dots [$(N,S) \rightarrow
(N+1,S')$] provides an important contribution to the conductance. It
is then sufficient to consider the two ground-state manifolds $(N,S)$
and $(N+1,S')$ which are respectively $(2S+1)$- and $(2S'+1)$-fold
degenerate. In this case, there is only one equation
(\ref{idb_simplified}) and a solution can always be
found;\cite{spin-degeneracy} the two variables $\Psi_S^{(N)}$ and
$\Psi_{S'}^{(N+1)}$ are uniquely determined by
\begin{equation}\label{gs-psi} 
\Psi_{S'}^{(N+1)} =
\Psi_S^{(N)} - \frac{\eta_{\rm l} \tilde{\Gamma}_{S,S'}^{\rm l}
- \eta_{\rm r} \tilde{\Gamma}_{S,S'}^{\rm r}} {\tilde{\Gamma}_{S,S'}^{\rm l} 
+ \tilde{\Gamma}_{S,S'}^{\rm r}}
\end{equation}
and by the normalization condition~(\ref{normalization1}). The linear
conductance then yields
\begin{widetext}
\begin{equation}\label{gs-conductance}
G_{\rm el} = \frac{e^2}{\hbar kT}
\frac{[2\,{\rm max}(S,S')+1]
f(\varepsilon_{S'}^{(N+1)}-\varepsilon_S^{(N)} - {\tilde \epsilon}_{\rm F})}
{(2S+1)+(2S'+1)e^{-\beta(\varepsilon_{S'}^{(N+1)} - \varepsilon_S^{(N)} - 
{\tilde \epsilon}_{\rm F})}}
\frac{\Gamma_{S,S'}^{\rm l} \Gamma_{S,S'}^{\rm r}}
{\Gamma_{S,S'}^{\rm l} + \Gamma_{S,S'}^{\rm r}}
\,,
\end{equation}
\end{widetext}
where the widths $\Gamma_{S,S'}^{\rm l,r}$ correspond to the
maximally-projected spin states with $M=S$ and $M'=S'$
[cf.~Eqs.~(\ref{maxproj}) and
(\ref{reduced-width-maxproj})]. According to
Eq.~(\ref{gs-conductance}), the functional form of the conductance
(versus the effective Fermi energy) does not depend on tunneling
widths. In particular, $G_{\rm el}$ is maximized when the effective
Fermi energy ${\tilde \epsilon}_{\rm F} \equiv e\zeta V_{\rm g} +
\epsilon_{\rm F}$ is tuned to
\begin{equation}
\label{g_max}
{\tilde \epsilon}_{\rm F,max} = \varepsilon_{S'}^{(N+1)} -
\varepsilon_S^{(N)} - \frac{kT}{2} \ln \left(\frac{2S'+1}{2S+1}\right)\;.
\end{equation}

\subsection{Ground-state--to--excited-states transitions}

We consider only a single state in either the $N$- or the
$(N+1)$-electron system and allow for any number of states in the
other system. This is useful at low temperatures, when for an even
number of electrons the lowest states in the spin sectors of $S=0$ and
$S=1$ can be close in energy. The approximation of the previous
paragraph may then be poor even at very low temperatures. For example,
one might consider the transitions $(N,S=1/2) \rightarrow
(N+1,\alpha',S')$ with $S'=0$ and $1$. The variables associated to the
$(N+1)$-electrons state are chosen as
\begin{equation}\label{psi-solution} \Psi_{\alpha' S'}^{(N+1)} =
\Psi_S^{(N)} - \frac{\eta_{\rm l} \tilde{\Gamma}_{S,\alpha'
    S'}^{\rm l} - \eta_{\rm r} \tilde{\Gamma}_{S,\alpha' S'}^{\rm
    r}} {\tilde{\Gamma}_{S,\alpha' S'}^{\rm l} +
  \tilde{\Gamma}_{S,\alpha' S'}^{\rm r}}\,,
\end{equation}
and the variable $\Psi_S^{(N)}$ is then fixed by the normalization
condition~(\ref{normalization1}). Clearly, Eqs.~(\ref{psi-solution})
solve Eqs.~(\ref{idb_simplified}) for all $\alpha'$ and the
conductance has the form (\ref{closed-conductance}).

\subsection{Systems with good orbital occupation numbers}

We define $n_\lambda = n_{\lambda +} + n_{\lambda -}$ to be the total
occupation of the (doubly-degenerate) single-particle orbital
$\lambda$, i.e. the sum of the number of spin-up and spin-down
particles in that orbital. Both $n_{\lambda +}$ and $n_{\lambda -}$
can take the value 0 or 1, hence $n_\lambda$ obtains the values 0, 1
or 2.  We will show below that Eqs.~(\ref{idb_simplified1}) have a
solution when all $n_\lambda$ are good quantum numbers. For a given
pair $(\alpha,S)$ and $(\alpha', S')$ only one term $\lambda =
\lambda_0$ in $\psi_\sigma^\dag({\rm \bf r}) = \sum_\lambda
\psi_\lambda ({\rm \bf r}) a^\dag_{\lambda \sigma}$ will then
contribute to the respective matrix element
\begin{eqnarray}\label{reduced-element}
(\alpha' S' \| \psi^\dag({\rm \bf r}) \| \alpha S) &=& 
\sum_\lambda \psi_\lambda ({\rm \bf r}) 
(\alpha' S' \| a_\lambda^\dag \| \alpha S)\nonumber\\
&=& \psi_{\lambda_0}({\rm \bf r}) (\alpha' S' \| a_{\lambda_0}^\dag 
\| \alpha S)\,.
\end{eqnarray}
Clearly, if more than one orbital $\lambda$ contribute to the sum in
Eq.~(\ref{reduced-element}), then the occupations cannot be good
quantum numbers in the final state $\alpha'$.  Since the dependence of
the reduced matrix element in Eq.~(\ref{reduced-element}) on the point
contact at ${\rm \bf r}$ is only through the wave function
$\psi_{\lambda_0}({\rm \bf r})$, we have
\begin{equation}
\frac{\eta_{\rm l} \tilde{\Gamma}_{\alpha S,\alpha' S'}^{\rm l} - 
\eta_{\rm r} \tilde{\Gamma}_{\alpha S,\alpha' S'}^{\rm r}} 
{\tilde{\Gamma}_{\alpha S, \alpha' S'}^{\rm l} + 
\tilde{\Gamma}_{\alpha S,\alpha' S'}^{\rm r}}
= \frac{\eta_{\rm l} \Gamma_{\lambda_0}^{\rm l} - \eta_{\rm r} 
\Gamma_{\lambda_0}^{\rm r}} {\Gamma_{\lambda_0}^{\rm l} + 
\Gamma_{\lambda_0}^{\rm r}}\;,
\end{equation}
where 
\begin{equation}
\Gamma_{\lambda}^{\rm l,r} =\Gamma_0^{\rm l,r}
|\psi_{\lambda}({\rm \bf r}_{\rm
l,r})|^2
\end{equation}
are the single-particle decay widths.

We now argue that Eqs.~(\ref{idb_simplified}) have a solution given by
\begin{eqnarray}\label{solution} 
\lefteqn{\Psi_{\alpha S}^{(m)} = -\sum_\lambda \frac{\eta_{\rm l} 
\Gamma_\lambda^{\rm l} - \eta_{\rm r} \Gamma_\lambda^{\rm r}}
{\Gamma_\lambda^{\rm l} + \Gamma_\lambda^{\rm r}} n_\lambda(\alpha,S)}
\nonumber\\
&&\hspace{30mm} \textrm{for } m=N,N+1\;,
\end{eqnarray}
where $n_\lambda(\alpha,S)$ are the single-particle level occupation
numbers of the states with quantum numbers $\alpha$ and $S$ [note that
the label $\alpha$ includes all the good quantum numbers
$\{n_\lambda\}=(n_1,n_2,\ldots)$ plus any additional quantum numbers
required to distinguish between states with the same occupations
$\{n_\lambda\}$ and spin $S$]. Since two manifolds of states
$(\alpha,S)$ and $(\alpha',S')$ can only be connected by changing the
occupation of a particular orbital $\lambda_0$ by a single particle,
we have
\begin{eqnarray}
\lefteqn{\Psi_{\alpha S}^{(N)} - \Psi_{\alpha' S'}^{(N+1)}}\nonumber\\
 &&= \sum_\lambda
\frac{\eta_{\rm l} \Gamma_\lambda^{\rm l} - \eta_{\rm r} 
\Gamma_\lambda^{\rm r}} {\Gamma_\lambda^{\rm l} + \Gamma_\lambda^{\rm r}}
[n_\lambda(\alpha,S) - n_\lambda(\alpha',S')]\nonumber\\
&&= \frac{\eta_{\rm l} \Gamma_{\lambda_0}^{\rm l} - 
\eta_{\rm r} \Gamma_{\lambda_0}^{\rm r}} {\Gamma_{\lambda_0}^{\rm l} + 
\Gamma_{\lambda_0}^{\rm r}} \;.
\end{eqnarray}
Thus Eqs.~(\ref{idb_simplified1}) can be indeed satisfied by (\ref{solution}).
For the conductance, we obtain the expression
\begin{eqnarray}\label{conductance3}
G &=& \frac{e^2}{\hbar kT} \sum_{\alpha S \atop \alpha' S'} {\tilde
P}_{\alpha S}^{(N)} f(\varepsilon_{\alpha S, \alpha' S'})\nonumber\\
&& \hspace{10mm} \times |(\alpha' S'
\| a_{\lambda_0}^\dag \| \alpha S)|^2 \frac{\Gamma_{\lambda_0}^{\rm l} 
\Gamma_{\lambda_0}^{\rm
r}} {\Gamma_{\lambda_0}^{\rm l} + \Gamma_{\lambda_0}^{\rm r}}\,,
\end{eqnarray}
where the orbital ${\lambda_0}$ depends on both the manifolds
$(\alpha,S)$ and $(\alpha',S')$. Of course, only manifolds that are
connected by the addition of an electron to a single-particle orbital
contribute to the sum in Eq.~(\ref{conductance3}); i.e.\ the
occupations $\{n_\lambda(\alpha',S')\}$ of the manifold $(\alpha',S')$
can only differ for one orbital and by one unit from the occupations
$\{n_\lambda(\alpha,S)\}$ of the manifold $(\alpha,S)$. An expression
similar to (\ref{conductance3}) (but including explicit sums over the
magnetic quantum numbers) was discussed in Ref.~\onlinecite{usaj01}.

An important example of a Hamiltonian in the class discussed here is
the so-called universal Hamiltonian in the limit of the Thouless
conductance $g_{\rm T}\to \infty$,
\begin{equation}\label{universal}
{\hat H} = \sum_{\lambda \sigma}\epsilon_\lambda 
a^\dagger_{\lambda \sigma}a_{\lambda \sigma} + {e^2\over 2 C} {\hat n}^2 
- J_S {\bf{\hat S}}^2\,,
\end{equation}
where ${\bf{\hat S}} = \sum_\lambda {\bf{\hat S}}_\lambda$ with
${\bf{\hat S}}_\lambda = \sum_{\sigma \sigma'} a^\dagger_{\lambda
\sigma}{\bm \sigma}_{\sigma \sigma'}a_{\lambda \sigma'}$ is the
total-spin operator of the dot and ${\bm \sigma}$ is the vector of the
three $2 \times 2$ Pauli matrices. Since $[\hat n_\lambda,{\bf{\hat
S}}_\lambda]=0$, the Hamiltonian (\ref{universal}) is invariant under
spin rotations and characterized by the good quantum numbers
$\{n_\lambda\}, S,$ and $M$. The conductance through a weakly-coupled
dot with a Hamiltonian (\ref{universal}) is then given by
Eq.~(\ref{conductance3}).

\subsection{\label{sec:KG} Systems with degenerate ground states}

In this subsection, we consider the case of low temperatures, when
only transitions between the ground states with $N$ and $N+1$
electrons in the dot are allowed, and these ground states are $M_N$-
and $M_{N+1}$-fold degenerate, respectively. This case is more general
than the similar case of Section~\ref{subsec:lowtemp}, since the
degeneracy is not necessarily the usual spin degeneracy, and the
transition widths $\Gamma^{\rm l,r}_{ij}$ can be arbitrary and are not
necessarily related by a Wigner-Eckart theorem. Although we cannot
write a closed solution for $G_{\rm el}$ in this more general case,
the degeneracy of the levels participating in the transport process
allows for significant simplification of the expression for $G_{\rm
el}$ [as compared with the general expression given by
Eqs.~(\ref{trafo_rate_eqs}), (\ref{normalization1}), and
(\ref{elastic_conductance})].

The level degeneracy allows us to reduce the rate equations
(\ref{trafo_rate_eqs}) to
\begin{subequations}
\label{deg_trafo_rate_eqs}
\begin{eqnarray}
\sum_j [(\Gamma_{ij}^{\rm l} + \Gamma_{ij}^{\rm r})
(\Psi_j^{(N+1)}\!-\!\Psi_i^{(N)})\hspace{3mm}&&
\nonumber\\
 + (\eta_{\rm l}
\Gamma_{ij}^{\rm l} - \eta_{\rm r} 
\Gamma_{ij}^{\rm r})]=0&&
\ \textrm{for all } i,
\label{deg_trafo_rate_eqns_1} 
\\[3mm] 
\sum_i 
[(\Gamma_{ij}^{\rm l} + \Gamma_{ij}^{\rm r}) (\Psi_j^{(N+1)}\!-\!
\Psi_i^{(N)})\hspace{3mm} &&
\nonumber\\
+ (\eta_{\rm l} \Gamma_{ij}^{\rm l} - \eta_{\rm r}
\Gamma_{ij}^{\rm r})]=0&&
\ \textrm{for all } j.
\label{deg_trafo_rate_eqns_2}
\end{eqnarray}
\end{subequations}
As mentioned in Section~\ref{section:linear-response}, only
$M_N+M_{N+1}-1$ of these $M_N+M_{N+1}$ equations are linearly
independent. The equation needed to determine a unique solution is
given by Eq.~(\ref{normalization1}).  In the special case discussed
here, the equilibrium probabilities for $N$ and $N+1$ electrons are
independent of the states $i$ and $j$, respectively, and
Eq.~(\ref{normalization1}) simplifies to
\begin{equation}\label{deg_normalization1}
 {\tilde P}^{(N)} \sum_i   \Psi_i^{(N)}   + 
{\tilde P}^{(N+1)}\sum_j  \Psi_j^{(N+1)} = 0\;,
\end{equation}
where we have defined $\tilde{P}^{(N)} = \tilde{P}_i^{(N)}$ for all $i$
and $\tilde{P}^{(N+1)} = \tilde{P}_j^{(N+1)}$ for all $j$. In particular
\begin{equation}
\label{re}
\tilde{P}^{(N)}=\frac{1}{M_N + M_{N+1} 
e^{-\beta(\varepsilon^{(N+1)} - \varepsilon^{(N)} - 
\tilde{\epsilon}_{\rm F})}}\;,
\end{equation}
 where $\varepsilon^{(N)} =
\varepsilon^{(N)}_i$ for all $i$ and $\varepsilon^{(N+1)} =
\varepsilon^{(N+1)}_j$ for all $j$. 

Using Eq.~(\ref{re}), the general expression
(\ref{elastic_conductance}) for the conductance reduces to
\begin{eqnarray}
\label{deg_G}
G_{\rm el} &=& \frac{e^2}{\hbar kT}
\frac{f(\varepsilon^{(N+1)} - \varepsilon^{(N)} - \tilde{\epsilon}_{\rm F})}
{M_N + M_{N+1} e^{-\beta(\varepsilon^{(N+1)} - \varepsilon^{(N)} - 
\tilde{\epsilon}_{\rm F})}}\nonumber\\
&&\hspace{7mm}\times
\sum_{ij} (\Psi_i^{(N)} - \Psi_j^{(N+1)} - \eta_{\rm l}) \Gamma_{ij}^{\rm l}\,.
\end{eqnarray}
Here, $\Psi_i^{(N)}$ and $\Psi_j^{(N+1)}$ are a solution to
Eqs.~(\ref{deg_trafo_rate_eqs}).  These equations do not depend on the
temperature, and we can therefore choose a private solution
$\tilde\Psi_i^{(N)}$ and $\tilde\Psi_j^{(N+1)}$ that is
temperature-independent and determined solely by the tunneling
widths. The general solution of Eqs.~(\ref{deg_trafo_rate_eqs}) is
given by $\Psi_i^{(N)}=\tilde\Psi_i^{(N)}+c$ and
$\Psi_j^{(N+1)}=\tilde\Psi_j^{(N+1)}+c$, where $c$ is a constant. The
solution that satisfies the normalization condition
(\ref{deg_normalization1}) is the one with $c=- ({\tilde P}^{(N)}
\sum_i \tilde\Psi_i^{(N)} + {\tilde P}^{(N+1)}\sum_j
\tilde\Psi_j^{(N+1)})$. While this constant $c$ depends on
temperature, it drops out in the final expression for the conductance,
thus making the sum on the r.h.s.\ of Eq.~(\ref{deg_G})
temperature-independent.

The conductance in Eq.~(\ref{deg_G}) factorizes into two
contributions. The sum over $i$ and $j$ is completely determined by
the tunneling widths $\Gamma^{\rm l,r}_{ij}$ and is independent of
temperature and the effective Fermi energy (it is also independent of
$\eta_{\rm l}$ and $\eta_{\rm r}$ as is shown in the Appendix). The
prefactor of the sum on the r.h.s.\ of Eq.~(\ref{deg_G}) does not
depend on the tunneling widths $\Gamma_{ij}^{\rm l,r}$ and contains
the full dependence on the temperature and Fermi energy. This
prefactor determines the functional dependence of the conductance peak
on the gate voltage.  Similar to Eq.~(\ref{g_max}) in
Section~\ref{subsec:lowtemp}, the maximum of $G_{\rm el}$ is attained
when the effective Fermi energy is tuned to
\begin{equation}
\label{Vgmax}
\tilde{\epsilon}_{\rm F,max} = \varepsilon^{(N+1)} - \varepsilon^{(N)} 
- \frac{kT}{2}\ln\left(\frac{M_{N+1}}{M_N}\right)\;.
\end{equation}

Eqs.~(\ref{deg_trafo_rate_eqs}) need to be solved numerically. An exception is the case discussed in Section~\ref{subsec:lowtemp},  where the dot has spin-rotation symmetry
and the degeneracy of the ground state corresponds to the various
values of the spin projection $M$. In this case  Eqs.~(\ref{deg_trafo_rate_eqs})
can be solved in closed form [see Eq.~(\ref{gs-psi})], and the linear conductance (\ref{deg_G}) reduces then to
Eq.~(\ref{gs-conductance}).

\section{The rapid-thermalization limit}\label{section:inelastic-scattering}

The above derivation of the linear conductance has been done under the
assumption of dominantly elastic scattering, which is a good
approximation at sufficiently low temperatures. Although the exact
temperature dependence of inelastic scattering events is not well
understood, they should become more relevant with increasing
temperature, such that their width $\Gamma_{\rm in}$ will eventually
be of comparable size to $\Gamma_{\rm el}$. We therefore 
consider in this Section the rapid-thermalization limit $\Gamma_{\rm
in} \gg \Gamma_{\rm el}$.

We denote by $P(N)$ the probability of the dot to be in a $N$-electron
state, and by $P(i|N)$ the conditional probability of the dot to be in
a particular many-body state $i$ given the dot is occupied by $N$
electrons. In general
\begin{equation}
P_i^{(N)} = P(i|N) P(N)\;.
\label{cond_prob}
\end{equation}
In the rapid-thermalization limit, the conditional probabilities
$P(i|N)$ and $P(j|N+1)$ are always given by their thermal equilibrium
values $\tilde P(i|N)$ and $\tilde P(j|N+1)$, respectively. It is then
only the probabilities $P(N)$ and $P(N+1)$ that obtain non-equilibrium
values by a finite bias voltage $V$.  These probabilities satisfy the
following rate equations:
\begin{widetext}
\begin{subequations}
\label{inelastic-rate-eqs}
\begin{eqnarray}
\frac{\partial P(N)}{\partial t} & = & P(N\!+\!1) \sum_{i,j} \tilde P(j|N\!+\!1) 
[(1\!-\!f_{ij}^{\rm l}) \Gamma_{ij}^{\rm l} +
(1\!-\!f_{ij}^{\rm r}) \Gamma_{ij}^{\rm r}]
- P(N) 
\sum_{i,j} \tilde P(i|N) [f_{ij}^{\rm l} \Gamma_{ij}^{\rm l} +
        f_{ij}^{\rm r} \Gamma_{ij}^{\rm r}] \;,
\\
\frac{\partial P(N\!+\!1)}{\partial t} & = &  P(N) \sum_{i,j} \tilde P(i|N) 
[f_{ij}^{\rm l} \Gamma_{ij}^{\rm l} +
        f_{ij}^{\rm r} \Gamma_{ij}^{\rm r}]
- P(N\!+\!1) \sum_{i,j} \tilde P(j|N\!+\!1) [(1\!-\!f_{ij}^{\rm l}) 
\Gamma_{ij}^{\rm l} + (1\!-\!f_{ij}^{\rm r}) \Gamma_{ij}^{\rm r}] \;.
\end{eqnarray}
\end{subequations}
\end{widetext}

As in the elastic case, we are interested in stationary solutions
$\partial P(N)/\partial t = \partial P(N+1)/\partial t=0$, where the
rate at which electrons tunnel onto the dot ($N \rightarrow N+1$) is
equal to the rate at which electrons tunnel off the dot ($N+1
\rightarrow N$). For a small voltage $V$, we can expand these
probabilities around their equilibrium values $\tilde P(N)$ and
$\tilde P(N+1)$ to first order in $V$,
\begin{subequations}
\label{prob_exp_inelast}
\begin{eqnarray}
P(N)\!&\!=\!&\!{\tilde P}(N)[1+eV\beta \Phi^{(N)}]\,,
\label{prob_n_exp}
\\
P(N+1)\!&\!=\!&\!{\tilde P}(N+1)[1+eV\beta \Phi^{(N+1)}]\,,
\label{prob_np1_exp}
\end{eqnarray}
\end{subequations}
where $\Phi^{(N)}$ and $\Phi^{(N+1)}$ are unknown variables. We
observe that while in the elastic limit there were $M_N + M_{N+1}$
unknown variables, in the inelastic limit we are left with a
considerably simpler situation of only two unknown variables.
Consequently, an explicit expression for the conductance can always be
given in the inelastic limit.

In analogy to the elastic case, we proceed by substituting linear
expansions in Eqs.~(\ref{fermi_dirac_exp}) and
(\ref{prob_exp_inelast}) into Eqs.~(\ref{inelastic-rate-eqs}), and
imposing stationarity. Collecting the terms that are independent of
the bias voltage $V$, we obtain the usual detailed-balance equation at
equilibrium.  Keeping only terms linear in $V$, we arrive at
\begin{widetext}
\begin{eqnarray}
\lefteqn{ {\tilde P}(N+1) \sum_{ij} \tilde P(j |N+1) [\beta (1-f_{ij})(\Gamma_{ij}^{
\rm l} + \Gamma_{ij}^{\rm r}) \Phi^{(N+1)} -
f_{ij}'(\eta_{\rm l} \Gamma_{ij}^{\rm l} - \eta_{\rm r} 
\Gamma_{ij}^{\rm r})]}
\nonumber\\
&& - {\tilde P}(N)
\sum_{ij} \tilde P(i|N)[\beta f_{ij}(\Gamma_{ij}^{\rm l} + \Gamma_{ij}^{\rm r}) 
\Phi^{(N)} + 
f_{ij}' (\eta_{\rm l} \Gamma_{ij}^{\rm l} - \eta_{\rm r} 
\Gamma_{ij}^{\rm r})] =0 \;.
\label{lin_inelastic}
\end{eqnarray}
\end{widetext}
Using the equilibrium relations $\tilde P(N) \tilde P(i|N) =\tilde
P^{(N)}_i$ and $\tilde P(N+1) \tilde P(j|N+1) =\tilde P^{(N+1)}_j$,
together with Eqs.~(\ref{equi_prob_n_np1}), (\ref{fermi_dirac_rel1}),
and (\ref{fermi_dirac_rel2}), we can rewrite Eq.~(\ref{lin_inelastic})
in the form
\begin{eqnarray}
\sum_{ij} \tilde P_i^{(N)} f_{ij} \Big[ (\Phi^{(N+1)} - \Phi^{(N)})
(\Gamma_{ij}^{\rm l} + \Gamma_{ij}^{\rm r}) && \nonumber\\
 - (\eta_{\rm l}\Gamma_{ij}^{\rm l} - \eta_{\rm r}\Gamma_{ij}^{\rm r}) 
\Big] &\!=\!&0\,.
\label{lin_inelastic1}
\end{eqnarray}
The solution for $\Phi^{(N)}$ and $\Phi^{(N+1)}$ (up to an arbitrary
additive constant) is then given by
\begin{equation}
\Phi^{(N+1)} - \Phi^{(N)} = 
\frac{\sum_{ij} P_i^{(N)} f_{ij}
(\eta_{\rm l} \Gamma_{ij}^{\rm l} - \eta_{\rm r} \Gamma_{ij}^{\rm r})}
{\sum_{ij} P_i^{(N)} f_{ij} (\Gamma_{ij}^{\rm l} + \Gamma_{ij}^{\rm r})}\,.
\label{solution_inelastic}
\end{equation}
The additive constant is determined by the normalization condition
$\tilde P(N) \Phi^{(N)} + \tilde P(N+1) \Phi^{(N+1)} =0$.

To find the linear conductance $G_{\rm therm}$ in the limit of
rapid-thermalization, we substitute the expansions in
Eqs.~(\ref{fermi_dirac_exp}) and (\ref{prob_exp_inelast}) in the
general expression for the linear-response current in
Eq.~(\ref{current}). We obtain
\begin{equation}
G_{\rm therm}\! =\! \frac{e^2}{\hbar kT} \sum_{ij} {\tilde P}_i^{(N)} f_{ij}
(\Phi^{(N)}\!-\! \Phi^{(N+1)}\!-\! \eta_{\rm l}) \Gamma_{ij}^{\rm l}.
\end{equation}
With the solution in Eq.~(\ref{solution_inelastic}), we obtain a
closed expression for the rapid-thermalization conductance
\begin{equation}
G_{\rm therm}\! =\! \frac{e^2}{\hbar kT}
\frac{\left(\sum_{ij} {\tilde P}_i^{(N)}\! f_{ij} \Gamma_{ij}^{\rm l}\right)\!\!
\left(\sum_{kl} {\tilde P}_k^{(N)}\! f_{kl} \Gamma_{ij}^{\rm r}\right)}
{\sum_{rs} {\tilde P}_r^{(N)} f_{rs} 
(\Gamma_{rs}^{\rm l} + \Gamma_{rs}^{\rm r})}.
\label{conductance_inelastic}
\end{equation}
As in the elastic case, we can exploit the spin symmetries to perform
explicitly the summation over the spin projections $M$ and $M'$,
\begin{equation}
\sum_{ij} {\tilde P}_i^{(N)} f_{ij} \Gamma_{ij}^{\rm l,r}
= \sum_{\alpha S \alpha' S'} {\tilde P}_{\alpha S}^{(N)}
f(\varepsilon_{\alpha' S',\alpha S}) 
{\tilde \Gamma}_{\alpha S, \alpha' S'}^{\rm l,r}\,.
\end{equation}
This yields as a final result for the conductance in the case of
strong inelastic scattering
\begin{widetext}
\begin{equation}
G_{\rm therm} = \frac{e^2}{\hbar kT} \frac{\left( \sum\limits_{\alpha
S \alpha' S'} {\tilde P}^{(N)}_{\alpha S} f(\varepsilon_{\alpha S,
\alpha' S'}) {\tilde \Gamma}_{\alpha S \alpha' S'}^{\rm l} \right)
\left( \sum\limits_{\alpha S \alpha' S'} {\tilde P}^{(N)}_{\alpha S}
f(\varepsilon_{\alpha S, \alpha' S'}) {\tilde \Gamma}_{\alpha S
\alpha' S'}^{\rm r} \right)} {\left( \sum\limits_{\alpha S \alpha' S'}
{\tilde P}^{(N)}_{\alpha S} f(\varepsilon_{\alpha S, \alpha' S'})
({\tilde \Gamma}_{\alpha S \alpha' S'}^{\rm l} + {\tilde
\Gamma}_{\alpha S \alpha' S'}^{\rm r} \right)}\,.
\end{equation}
\end{widetext}

\section{Summary and conclusion}\label{section:summary}

We define 
\begin{eqnarray}\label{average}
\langle\!\!\!\langle X_{ij}\rangle\!\!\!\rangle &\equiv& \sum_{ij} {\tilde
P}_i^{(N)} f(\varepsilon_{ij}) X_{ij}\nonumber\\
&=& \sum_{ij} {\tilde P}_j^{(N+1)}
[1- f(\varepsilon_{ij})] X_{ij}\,,
\end{eqnarray}
where $X_{ij}$ is a quantity that depends on the many-body states $i$
and $j$ of the $N$- and $(N+1)$-electron dots, respectively. The
equilibrium probabilities $\tilde P_i^{(N)}$ and $\tilde P_j^{(N+1)}$
to find the dot in states $i$ and $j$ are given by
Eqs.~(\ref{def_equi_prob}) and (\ref{partition}), while the energy
difference $\varepsilon_{ij}$ is defined in Eq.~(\ref{energy-diff}).

In the rapid-thermalization limit, the conductance of the almost-isolated
dot is given by
\begin{equation}\label{inelastic}
G_{\rm therm}= {e^2\over \hbar kT}{\langle\!\!\!\langle
\Gamma_{ij}^{\rm l}\rangle\!\!\!\rangle \langle\!\!\!\langle
\Gamma_{ij}^{\rm r}\rangle\!\!\!\rangle \over \langle\!\!\!\langle
\Gamma_{ij}^{\rm l} + \Gamma_{ij}^{\rm r} \rangle\!\!\!\rangle} \;,
\end{equation}
where $\Gamma_{ij}^{\rm l}$ ($\Gamma_{ij}^{\rm r}$) are the partial
transition widths between the states $i$ and $j$ involving the
tunneling of an electron from the left (right) lead into the dot.

In the elastic limit, a closed solution is not possible in the general
case. The conductance can be calculated by solving the linear
equations (\ref{trafo_rate_eqs}) for $\Psi_i^{(N)}$ and
$\Psi_j^{(N+1)}$ and then substituting in the expression
(\ref{elastic_conductance}). However, there are several important
cases where a closed solution is possible, including the
constant-exchange-interaction model (i.e.\ the universal Hamiltonian
in the limit of infinite Thouless conductance). In these cases
\begin{equation}\label{elastic}
G_{\rm el}= {e^2\over \hbar kT}\left\langle\!\!\!\left\langle
{\Gamma_{ij}^{\rm l} \Gamma_{ij}^{\rm r} \over \Gamma_{ij}^{\rm l} +
  \Gamma_{ij}^{\rm r}}\right\rangle\!\!\!\right\rangle 
\;.
\end{equation}

When the Hamiltonian is invariant under spin rotations, these
expressions can be simplified by carrying out explicitly the summation
over the magnetic quantum numbers. The many-body levels of the $N$ and
$(N+1)$-electron dots are now characterized by the quantum numbers
$\alpha, S, M$ and $\alpha', S', M'$ and the respective energies are
independent of $M$ and $M'$. Equations~(\ref{inelastic}) and
(\ref{elastic}) are now valid with the reduced widths
$\tilde\Gamma_{\alpha S \alpha' S'}^{\rm l,r}$ replacing the widths
$\tilde\Gamma_{\alpha S M \alpha' S' M'}^{\rm l,r}$, and the summation
in Eq.~(\ref{average}) carried over $\alpha S$ and $\alpha' S'$ only
(but not over $M$ and $M'$).

In conclusion, we have solved the rate equations and found the linear
conductance in the presence of interactions in the dot (beyond the
charging energy). In particular, we have taken into account the spin
degrees of freedom of the dot and we showed the simplifications that
occur when the dot's Hamiltonian is invariant under spin rotations.
Both the limits of dominantly elastic scattering and rapid
thermalization are discussed. This work generalizes the solutions of
Ref.~\onlinecite{beenakker91} that were derived in the limit of
non-interacting electrons (except for a constant charging energy).

\section*{Acknowledgments}

The work at Yale was supported in part by the Department of Energy
grant No.\ DE-FG-0291-ER-40608; the work at the University of Maryland
was supported by the US-ONR, ARDA, and DARPA; and the work at the
University of Minnesota was supported by NSF Grants No.\ DMR 97-31756,
DMR 02-37296, and EIA 02-10736. We thank S.~Malhotra for helpful discussions.

\section*{Appendix}

Here we show that the linear conductance in
Eq.~(\ref{elastic_conductance}) is independent of $\eta_{\rm l}$ and
$\eta_{\rm r}$, the fractions of the bias potential difference between
the dot and the corresponding leads. Defining the new variables
$\Phi_i^{(N)} \equiv \Psi_i^{(N)} - \eta_{\rm l}$, we can rewrite
Eqs.~(\ref{trafo_rate_eqs}) in the form
\begin{subequations}
\label{trafo_rate_eqs'}
\begin{eqnarray}
\sum_j f_{ij} [(\Gamma_{ij}^{\rm l} + \Gamma_{ij}^{\rm r})
(\Psi_j^{(N+1)}\!-\!\Phi_i^{(N)}) - \Gamma_{ij}^{\rm r})]=0&&\nonumber\\
\textrm{for all } i\,,\hspace{2mm}&&
\label{trafo_rate_eqns'_1} 
\end{eqnarray}
\begin{eqnarray}
\sum_i (1-f_{ij})
[(\Gamma_{ij}^{\rm l} + \Gamma_{ij}^{\rm r}) (\Psi_j^{(N+1)}\!-\!
\Phi_i^{(N)}) -
\Gamma_{ij}^{\rm r})]= 0&&\nonumber\\ 
\textrm{for all } j\,.\hspace{5mm}&&
\label{trafo_rate_eqns'_2}
\end{eqnarray}
\end{subequations}
The solution of Eqs.~(\ref{trafo_rate_eqs'}) is determined up to an
additive constant. Since $\eta_{\rm l}$ and $\eta_{\rm r}$ do not
appear explicitly in Eqs.~(\ref{trafo_rate_eqs'}), we can choose a
private solution $\tilde\Phi_i^{(N)}$ and $\tilde\Psi_j^{(N+1)}$ that
is independent of $\eta_{\rm l}$ and $\eta_{\rm r}$. The general
solution is then given by $\Phi_i^{(N)}=\tilde\Phi_i^{(N)}+c$ and
$\Psi_j^{(N+1)}=\tilde\Psi_j^{(N+1)}+c$, where $c$ is a constant.  In
particular, the normalization condition (\ref{normalization1}) can be
satisfied by choosing
\begin{equation}\label{constant-c}
c=-\!\left[\sum_i\! \tilde P_i^{(N)} (\tilde\Phi_i^{(N)}\!+\!\eta_{\rm l}) +
\sum_j\! \tilde P_j^{(N+1)} \tilde\Psi_j^{(N+1)}\right].
\end{equation} 
The constant $c$ depends on $\eta_{\rm l}$, but disappears in the
final expression for the conductance in
Eq.~(\ref{elastic_conductance}),
\begin{equation}\label{elastic_conductance1}
G_{\rm el} = \frac{e^2}{\hbar kT} \sum_{ij} {\tilde P}_i^{(N)} f_{ij} 
(\tilde\Phi_i^{(N)} \!-\! \tilde\Psi_j^{(N+1)}) \Gamma_{ij}^{\rm l}.
\end{equation}
Expression (\ref{elastic_conductance1}) for the conductance shows
clearly its independence of $\eta_{\rm l}$ (and $\eta_{\rm r}$).


\begin{thebibliography}{99}
\bibitem{spin-degeneracy} 
  Here we have already used the fact that the original
  $\mbox{(2S+1)}\mbox{(2S'+1)}$ equations
  (\ref{individual_detailed_balance}) with \mbox{$i\!=\!(\alpha,S,M)$}
  and \mbox{$j\!=\!(\alpha',S',M')$} can be reduced to a single
  equation because of the Wigner-Eckart theorem (\ref{Wigner-Eckart}).
\end{thebibliography}
\end{document}